\documentclass[journal,twoside,web]{ieeecolor}
\usepackage{generic}
\usepackage{cite}
\usepackage{amsmath,amssymb,amsfonts}
\usepackage{pgfplots}
\pgfplotsset{compat=1.18}
\usepackage{algorithmic}
\usepackage{graphicx}
\usepackage{algorithm,algorithmic}
\usepackage{hyperref}
\hypersetup{hidelinks=true}
\usepackage{textcomp}
\usepackage{subfiles}
\usepackage{multirow}
\usepackage{booktabs}
\usepackage{xcolor}
\newcommand{\rev}[1]{#1}
\usepackage{subfig}
\usepackage{tikz}
\usepackage{subcaption}
\usepackage{threeparttable}
\usepackage[switch]{lineno}
\usetikzlibrary{shapes, arrows, positioning}

\def\BibTeX{{\rm B\kern-.05em{\sc i\kern-.025em b}\kern-.08em
    T\kern-.1667em\lower.7ex\hbox{E}\kern-.125emX}}
\begin{document}

\title{Combining scEEG and PPG for reliable sleep staging}
\author{Jiawei Wang, Liang Xu, Shuntian Zheng, Yu Guan, Kaichen Wang, Ziqing Zhang, Chen Chen, \\Laurence T. Yang, \IEEEmembership{Fellow, IEEE}, Sai Gu, \rev{M. Jamal Deen,
\IEEEmembership{Life Fellow, IEEE}}
\thanks{Jiawei Wang, Liang Xu, Sai Gu are with the School of Engineering, The University of Warwick, Coventry, UK (e-mail: Davy.Wang@warwick.ac.uk; Liang.Xu.1@warwick.ac.uk; Sai.Gu@warwick.ac.uk).}
\thanks{Shuntian Zheng and Yu Guan are with the Department of Computer Science, The University of Warwick, Coventry, UK (e-mail: Shuntian.Zheng@warwick.ac.uk;Yu.Guan@warwick.ac.uk).}
\thanks{Kaichen Wang, Ziqing Zhang are with the College of Biomedical Engineering, Fudan University, Shanghai, China (e-mail:  23110720140@m.fudan.edu.cn; 25212030032@m.fudan.edu.cn).}
\thanks{Chen Chen is with the Human Phenome Institute, Fudan University, Shanghai, China
 (e-mail: chenchen\_fd@fudan.edu.cn)}
\thanks{Laurence T. Yang is with the School
of Computer and Artificial Intelligence, Zhengzhou University, Zhengzhou, China
 (e-mail: ltyang@gmail.com)}
\thanks{\rev{M. Jamal Deen is with the Department of Electrical and Computer Engineering and the School of Biomedical Engineering, McMaster University, Hamilton, ON L8S 4K1, Canada (e-mail: jamal@mcmaster.ca)}}
 }

\maketitle

\begingroup\renewcommand\thefootnote{}\footnotetext{This work has been submitted to the IEEE for possible publication. Copyright may be transferred without notice, after which this version may no longer be accessible.}\endgroup

\begin{abstract}
Reliable sleep staging remains challenging for lightweight modalities such as single-channel electroencephalography (scEEG) or photoplethysmography (PPG). scEEG offers direct measurement of cortical activity and serves as the foundation for sleep staging, yet exhibits limited performance on light sleep stages. PPG provides a low-cost complement
that captures autonomic signatures effective for detecting light sleep. However, prior PPG-based methods rely on full night recordings (8 - 10 hours) as input context, which is less practical to provide timely feedback for sleep intervention. In this work, we investigate scEEG-PPG fusion for 4-class sleep staging under short-window (30\,s - 30\,min) constraints. \rev{All signals are derived from polysomnography using a single mastoid-referenced EEG channel and finger pulse-oximetry PPG as proxies for wearable-compatible signals.} First, we evaluate the temporal context required for each modality, to better understand the relationship of sleep staging performance with respect to monitoring window. Second, we investigate three fusion strategies: score-level fusion, cross-attention fusion enabling feature-level interactions, and Mamba-enhanced fusion incorporating temporal context modeling. Third, we train and evaluate on the Multi-Ethnic Study of Atherosclerosis (MESA) dataset and perform cross-dataset validation on the Cleveland Family Study (CFS) and the Apnea, Bariatric surgery, and CPAP (ABC) datasets. The Mamba-enhanced fusion achieves the best performance on MESA (Cohen's Kappa $\kappa$ = 0.798, Acc = 86.9\%), with particularly notable improvement in light sleep classification (F1-score: 85.63\% vs. 77.76\%, recall: 82.85\% vs. 69.95\% for scEEG alone), and generalizes well to CFS and ABC datasets with different populations. These findings suggest that scEEG-PPG fusion is a promising approach for sleep monitoring, offering a pathway toward more accessible sleep health assessment. Source code of this project can be found at:
\href{https://github.com/DavyWJW/scEEG-PPG-Fusion}{https://github.com/DavyWJW/scEEG-PPG-Fusion}

\end{abstract}

\begin{IEEEkeywords}
Sleep stage classification, AI for healthcare, multimodal fusion

\end{IEEEkeywords}

\section{Introduction}
\begin{figure*}[!t]
\centering
\includegraphics[width=0.9\textwidth]{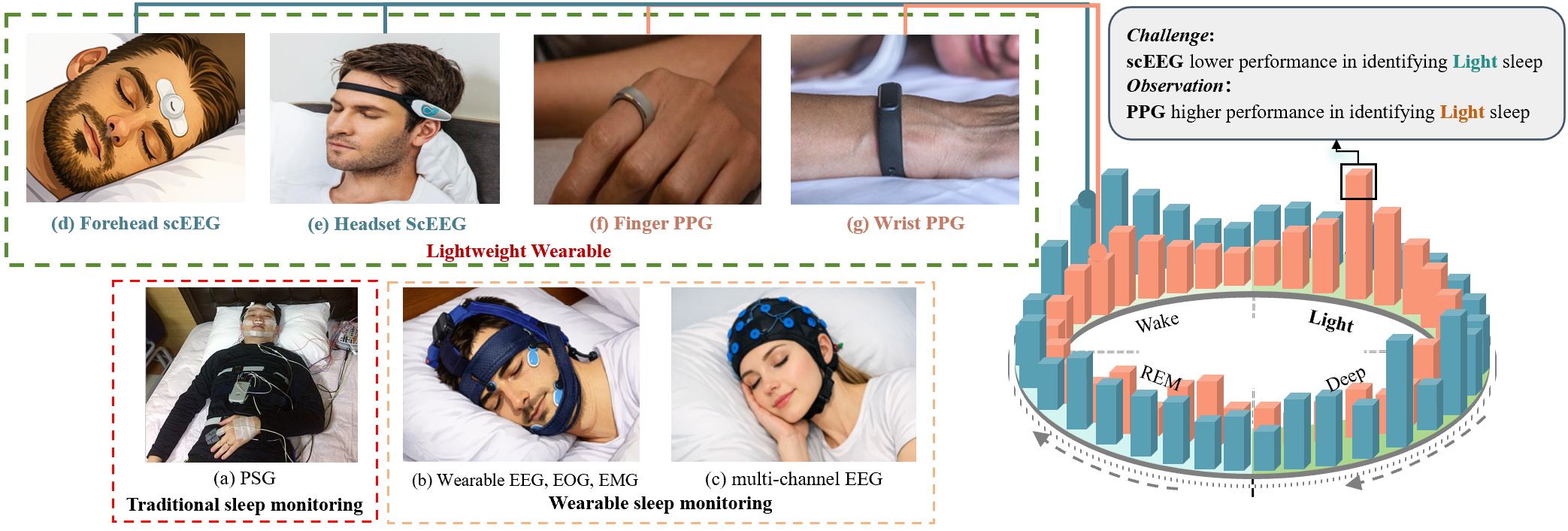}
\caption{Wearable sleep monitoring devices: (a) clinical gold-standard PSG \cite{wikipedia_polysomnography}, (b) multi-signal system integrating EEG, EOG, and EMG (Tosoo AG) \cite{eicher2024phase}, (c) multi-channel EEG system (Greentek GT Cap) \cite{greenteksensor_gtcap}, (d) forehead scEEG device (Neurovista Sleep Monitor) \cite{Neurovista2024}, (e) headset scEEG device (Brainlink Lite) \cite{abdal2025eeg}, (f) PPG-based smart ring (RingConn) \cite{wang2024will}, (g) PPG-based smart watch (Garmin) \cite{macdermott2019forensic}.}
\label{fig:wearable}
\end{figure*}

Sleep plays a vital role in maintaining human health, affecting cognitive performance, metabolic balance, and cardiovascular regulation \cite{ting2005disorders}. Accurate sleep staging, as defined by the American Academy of Sleep Medicine (AASM)~\cite{berry2012aasm}, involves classifying nocturnal recordings into wakefulness (Wake), rapid eye movement (REM), and non-REM stages (N1/N2/N3), and is essential for diagnosing sleep disorders and assessing overall sleep quality. Polysomnography (PSG), the clinical gold standard, relies primarily on multi-channel electroencephalography (EEG), electrooculography (EOG), electromyography (EMG), and electrocardiogram (ECG), etc.\cite{rundo2019polysomnography}.
As shown in Fig.~\ref{fig:wearable}(a), despite its reliability, the high cost, cumbersome setup, and dependence on professional scoring restrict the scalability of PSG for daily or home-based monitoring \cite{wikipedia_polysomnography}. For practical applications, a four-stage classification comprising Wake, REM, light sleep (N1+N2), and deep sleep (N3) is commonly adopted \cite{boe2019automating}.

With advances in wearable sensing and remote health monitoring
\cite{majumder2017wearable}, single-channel EEG devices have emerged as an
accessible alternative for portable sleep monitoring.
Fig.~\ref{fig:wearable}(b--e) shows representative devices: (b) multi-signal
systems integrating EEG, EOG, and EMG, (c) multi-channel EEG systems,
(d) forehead single-channel EEG (scEEG) devices and (e) headset scEEG devices.
Compared to PSG, these devices offer improved portability and ease of
deployment. Among them, scEEG represents the most lightweight solution for
home-based sleep monitoring, requiring only minimal electrode setup with
significantly lower cost and improved user comfort compared to multi-signal
and multi-channel systems \cite{eicher2024phase}, making it well suited to the
unobtrusive, long-term monitoring envisioned for smart-home and
ambient-assisted-living settings \cite{deen2015information}.
However, scEEG generally exhibits lower accuracy for detecting light sleep
stage \cite{salfi2026potential}.

This difficulty in light sleep classification reflects challenges acknowledged by AASM guidelines, which characterize light sleep (particularly stage N1) as a transitional state with subtle EEG features that can be easily confused with Wake or REM sleep \cite{berry2017aasm}. For scEEG, the limited spatial information further exacerbates this challenge, as characteristic features may not be adequately captured from a single recording site.

Research has demonstrated that different sleep stages are associated with distinct autonomic patterns, with the transition from Wake to light sleep showing pronounced shifts from sympathetic to parasympathetic dominance \cite{hedner2011sleep}. This suggests that autonomic signals may complement EEG particularly for light sleep identification. Photoplethysmography (PPG) is an optical technique that measures blood volume changes through light absorption \cite{allen2021photoplethysmography}, commonly deployed in smart rings or smart watches as shown in Fig.~\ref{fig:wearable}(f) and (g). PPG captures pulse morphology and autonomic dynamics during sleep transitions \cite{kotzen2022sleepppg}. While PPG-based sleep staging methods generally achieve lower overall accuracy than scEEG-based approaches due to the indirect nature of autonomic signals, PPG may provide complementary information where scEEG features are ambiguous, as schematically illustrated in Fig.~\ref{fig:wearable}, particularly during light sleep, based on prior empirical findings \cite{kotzen2022sleepppg, eldele2021attention}. Furthermore, as a lightweight and low-cost sensor, PPG is accessible for augmenting scEEG-based sleep monitoring \cite{de2019wearable}.

This motivates investigating to what extent PPG can complement scEEG for sleep staging. Unlike EEG which directly captures cortical sleep signatures, PPG relies on indirect autonomic nervous system modulation of heart rate variability and pulse waveform characteristics \cite{motin2023multi}. These cardiovascular dynamics evolve slowly and require extended observation periods to reveal discriminative patterns, which explains why state-of-the-art (SOTA) PPG methods process full-night recordings (8 - 10 h) as input context \cite{wang2025improving, kotzen2022sleepppg}. For lightweight applications, short analysis windows are essential. Whether PPG can retain sufficient discriminative power under such constraints, and how to effectively combine the distinct characteristics of scEEG and PPG, remain open questions. \rev{Although multimodal fusion has been studied for sleep staging, for example, by combining EEG with ECG~\cite{lyu2023novel} or EOG~\cite{yu2025casleepnet}, or by fusing movement and cardiac signals~\cite{zhai2020making, fonseca2020automatic}, these methods largely target full-night or clinical multi-signal recordings rather than the short-window scEEG and PPG pairing pursued here.}

To address these issues, this study presents an investigation of combining scEEG and PPG for sleep staging. Leveraging existing scEEG and PPG sleep staging algorithms \cite{eldele2021attention, wang2025improving}, we first evaluate performance across multiple window lengths (30 s - 30 min) to study the minimal temporal context required for reliable sleep monitoring. We then investigate three fusion strategies: score-level fusion, cross-attention fusion that enables bidirectional feature-level interactions between modalities, and Mamba-enhanced fusion that incorporates temporal context modeling through selective state space models \cite{gu2024mamba}. We train and evaluate our models on the Multi-Ethnic Study of Atherosclerosis (MESA) dataset \cite{chen2015racial}, a large-scale community-based sleep dataset. Furthermore, we evaluate generalizability through cross-dataset validation on the Cleveland Family Study (CFS) and the Apnea, Bariatric surgery, and CPAP (ABC) datasets \cite{redline1995familial,bakker2018gastric}, including different populations.

The main contributions of this work are summarized as follows:

\begin{itemize}
    \item To the best of our knowledge, this is the first study to investigate the fusion of scEEG and PPG for short-window sleep staging. We explore the feasibility of using PPG to complement scEEG for lightweight sleep monitoring, addressing the limitation of scEEG in light sleep classification through multimodal fusion.
    \item We conduct a short-window study comparing scEEG and PPG across window lengths from 30 s to 30 min, and identify that 3-minute windows provide an effective trade-off between performance and window length.
\item We investigate three fusion strategies (score-level, cross-attention, and Mamba-enhanced), demonstrating that Mamba-enhanced fusion achieves the best performance (Cohen's Kappa $\kappa$ = 0.798) on MESA dataset, with light sleep classification significantly improved (F1-score: 77.76\% $\rightarrow$ 85.63\%, recall: 69.95\% $\rightarrow$ 82.85\% for scEEG alone), and generalizes effectively to other datasets. 
\end{itemize}

\section{Methodology}

\subsection{Datasets}
\subsubsection{MESA Dataset}

\begin{figure*}[!t]
\centering
\includegraphics[width=\textwidth]{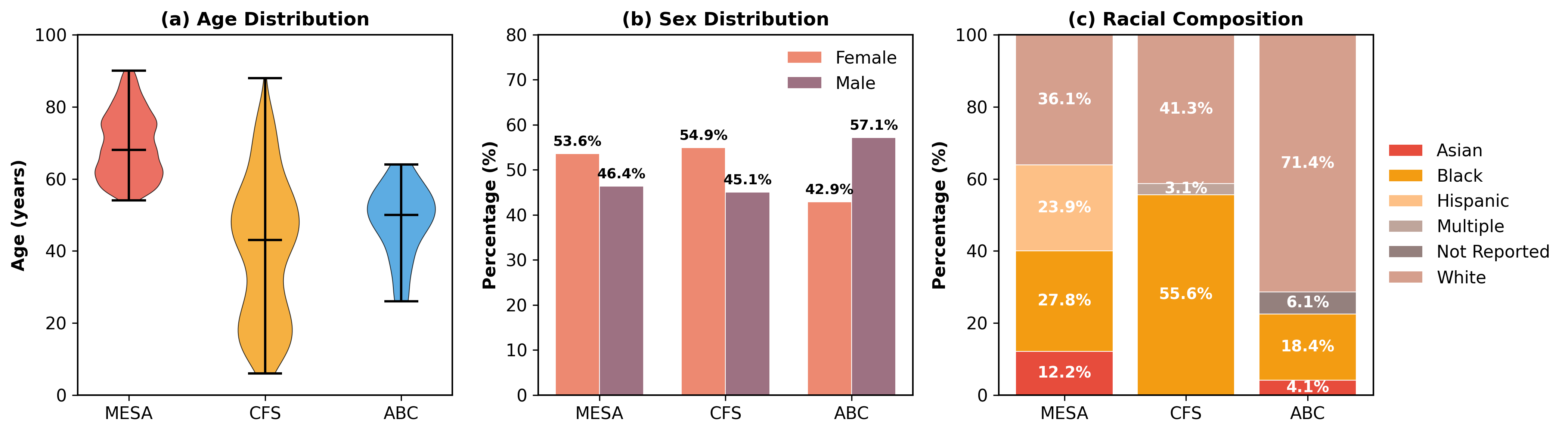}
\caption{Demographic characteristics of MESA, CFS and ABC datasets: (a) Age. (b) Sex distribution. (c) Racial composition highlighting MESA's ethnic diversity, CFS's predominantly Black and White population, and ABC's majority White cohort.}
\label{fig:demographics}
\end{figure*}

MESA is a National Heart, Lung, and Blood Institute (NHLBI)-sponsored multi-center longitudinal investigation of cardiovascular disease risk factors \cite{chung2023multi}. Between 2010 and 2012, 2,237 participants were enrolled in the MESA sleep ancillary study with full overnight unattended PSG. The cohort comprises ethnically diverse participants (38\% White, 28\% Black, 23\% Hispanic, and 11\% Chinese-American) aged 54 - 93 years, representing one of the largest and most diverse community-based sleep datasets available. Each PSG recording includes synchronized scEEG, PPG (from finger pulse oximetry) signals, and expert-annotated sleep stages following AASM guidelines\cite{chen2015racial}. After excluding recordings with missing or corrupted signals, 2,056 subjects are selected. Following prior work \cite{wang2025improving}, we use a predefined test set of 204 subjects, with the remaining 1,852 subjects split into training and validation sets.

\subsubsection{CFS Dataset}
CFS is the largest family-based study of sleep apnea worldwide, comprising 2,284 individuals (46\% African American) from 361 families studied over a 16-year period \cite{redline1995familial}. The study was designed to investigate genetic and non-genetic risk factors for sleep-disordered breathing. PSG recordings were collected during the fourth examination (2001-2006) at a general clinical research center, including EEG, EOG, EMG, ECG, PPG and respiratory signals with expert-scored sleep stages. For cross-dataset evaluation, we utilize 719 subjects, each contributing up to 10 hours of scEEG and PPG recordings. CFS presents distinct challenges for generalization due to differences in recording protocols, equipment, and demographic composition compared to MESA, including a broader age range and higher proportion of subjects with sleep-disordered breathing.

\subsubsection{ABC Dataset}
ABC study is a randomized clinical trial comparing bariatric surgery versus continuous positive airway pressure therapy for patients with class II obesity and severe obstructive sleep apnea (OSA) \cite{bakker2018gastric}. The dataset includes 49 subjects who underwent baseline PSG with synchronized scEEG and PPG recordings. ABC represents a significant clinical domain shift from MESA, as participants exhibit altered sleep architecture characterized by frequent arousals, reduced deep sleep proportion, and disrupted autonomic regulation due to severe OSA. This dataset enables evaluation of model robustness under challenging pathological conditions that are common in real-world clinical populations.

Fig.~\ref{fig:demographics} summarizes the demographic characteristics across all three datasets. MESA represents an older cohort (mean age 69.4 years) with substantial ethnic diversity, providing a robust foundation for model training. CFS spans a broader age range including younger participants (mean age 41.2 years) with predominantly Black (55.6\%) and White (41.3\%) populations. ABC comprises middle-aged adults (mean age 48.2 years) with obesity and severe OSA, predominantly White (71.4\%), representing a distinct clinical population for evaluating model robustness under pathological conditions.

\subsection{Signal Preprocessing}
\subsubsection{scEEG Preprocessing}

All three datasets employ mastoid-referenced EEG derivations recommended by AASM guidelines. We use a single EEG derivation from PSG as a proxy for wearable-compatible scEEG. Specifically, MESA provides C4-M1 (right central electrode referenced to left mastoid), while CFS and ABC utilize C3-M2 (left central electrode referenced to right mastoid). This setup isolates the modeling question under a single-channel constraint, while acknowledging that prospective on-device validation remains necessary (see Limitations in Section~\ref{sec:discussion}). Following the scEEG preprocessing procedure in \cite{eldele2021attention}, raw signals undergo bandpass filtering between 0.3 and 35 Hz and resampled to 100 Hz, yielding 3,000 samples per 30-second epoch. Continuous recordings are segmented into non-overlapping epochs aligned with expert annotations, with consecutive epochs concatenated for multi-epoch windows (1 - 30 minutes). Recording-level z-score normalization standardizes amplitude across subjects.

\subsubsection{PPG Preprocessing}
Following the PPG preprocessing procedure in \cite{wang2025improving}, for MESA, finger photoplethysmography signals undergo 8 Hz lowpass filtering using an 8th-order Chebyshev Type II filter, followed by resampling to 34.13 Hz (1,024 samples per 30-second epoch). Values exceeding ±3 standard deviations are clipped, and recording-level z-score normalization is applied. CFS and ABC follow identical preprocessing pipeline to ensure cross-dataset compatibility.

\begin{figure}[t]
    \centering
    \includegraphics[width=1\columnwidth]{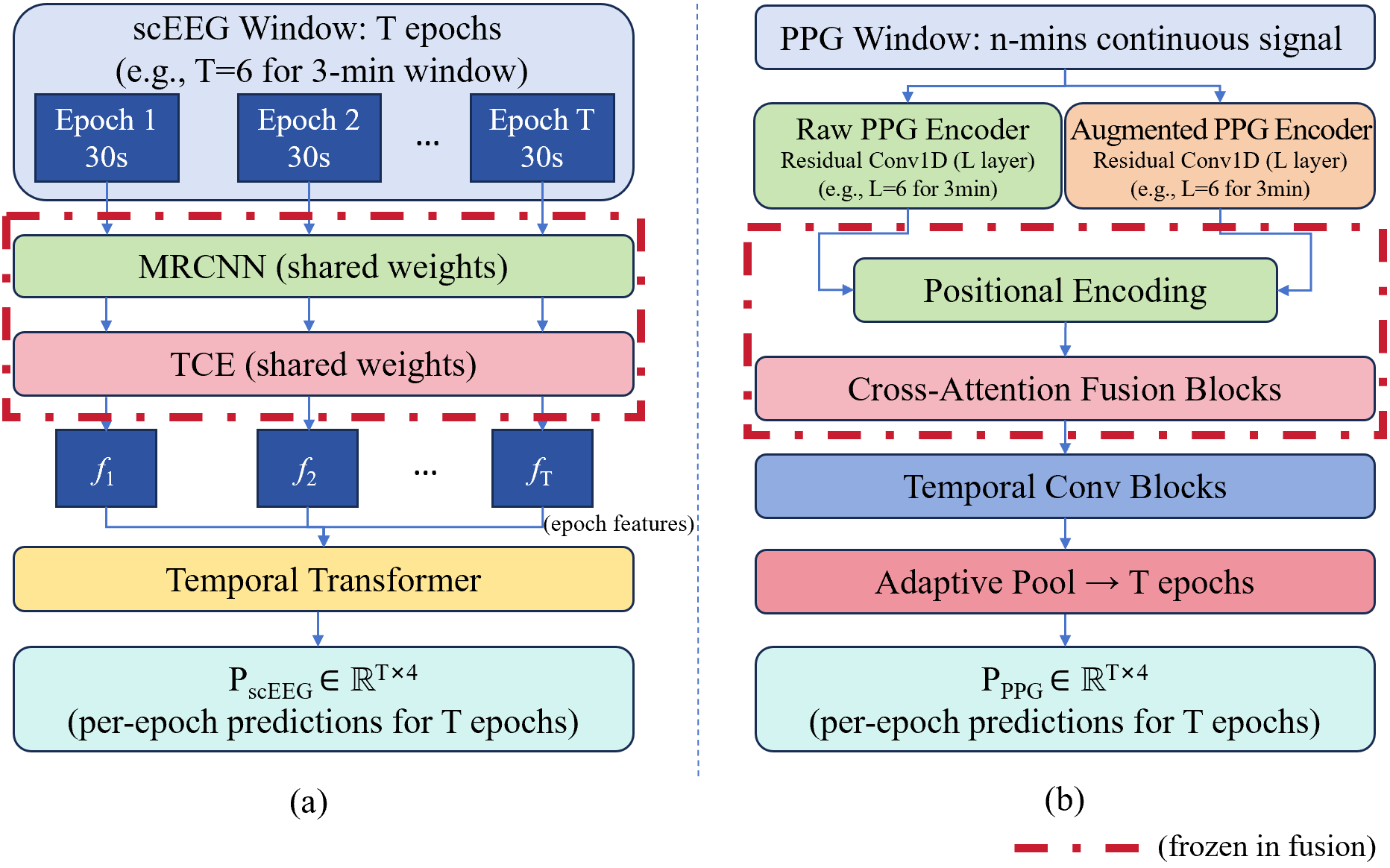}
    \caption{Architecture of baseline models: (a) scEEG model, (b) PPG model.}
    \label{fig:eeg}
\end{figure}

\begin{figure*}[t]
    \centering
    \includegraphics[width=\textwidth]{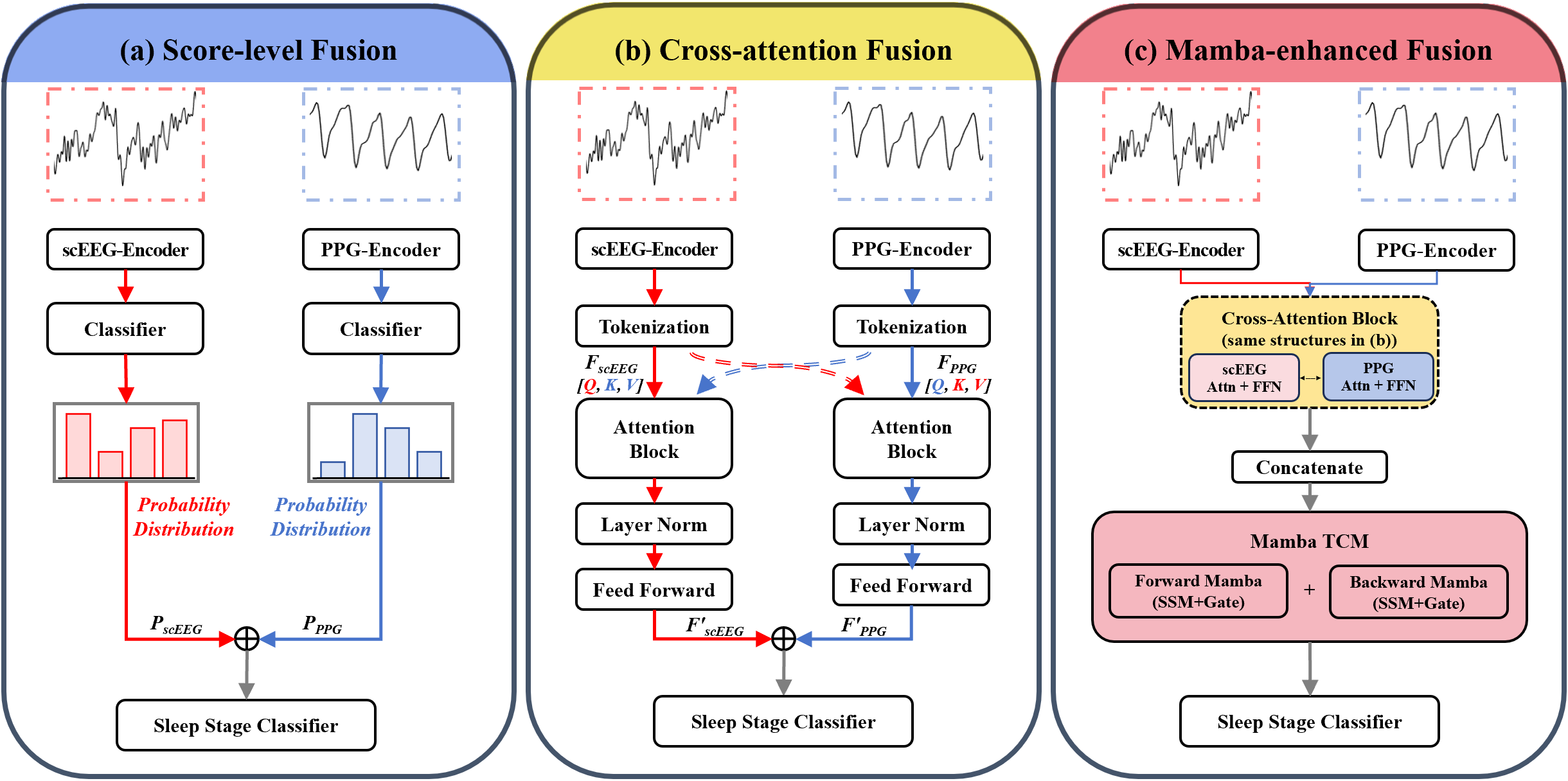}
    \caption{Overview of three fusion strategies: (a) score-level fusion, (b) cross-attention fusion, and (c) Mamba-enhanced fusion.}
    \label{fig:fusion_architectures}
\end{figure*}

\subsection{Model Architecture}
\subsubsection{scEEG-based model}
We adopt the AttnSleep architecture \cite{eldele2021attention}, a well-validated model for scEEG sleep staging, which employs a Multi-Resolution Convolutional Neural Network (MRCNN) to extract hierarchical temporal features, followed by a Transformer-based temporal context encoder (TCE) with multi-head self-attention to capture long-range dependencies, as illustrated in Fig.~\ref{fig:eeg}(a). To handle multi-epoch windows, we extend AttnSleep by concatenating epoch-level features and feeding them into a temporal transformer (fixed 2 layers, 256 dimensions) that models cross-epoch dependencies. The model size remains constant across window lengths, outputting $\mathbf{P}_{\mathrm{scEEG}} \in \mathbb{R}^{T \times 4}$, where $T$ is the number of epochs and 4 corresponds to the sleep stage classes (Wake, Light, Deep, REM).

\subsubsection{PPG-based model}
We adopt a dual-stream architecture processing both raw PPG and augmented PPG through parallel residual convolutional encoders \cite{wang2025improving}, as illustrated in  Fig.~\ref{fig:eeg}(b). Cross-attention fusion blocks with bidirectional multi-head attention enable information exchange between streams, followed by temporal convolution blocks for sequence modeling. An adaptive weighting module dynamically balances the contributions of the two streams. The model outputs $\mathbf{P}_{\mathrm{PPG}} \in \mathbb{R}^{T \times 4}$.

\subsubsection{scEEG-PPG fusion}
Having established separate encoders for scEEG and PPG, we now describe how to effectively combine their complementary representations. The fusion module aims to leverage the strengths of both modalities: scEEG excels at distinguishing deep sleep through delta wave patterns \cite{lechat2022novel}, whereas PPG provides robust features for light sleep detection through heart rate variability \cite{kim2022autonomic}. We study scEEG and PPG fusion strategies to improve sleep stage classification.
Both modalities are processed using aligned short windows ranging from 30 seconds to 30 minutes. In this work, we study fusion strategies, as illustrated in Fig.~\ref{fig:fusion_architectures}: score-level fusion, cross-attention fusion, and Mamba-enhanced fusion.

\textbf{(a) Score-level fusion:} As illustrated in Fig.~\ref{fig:fusion_architectures}(a), score-level fusion combines the probability outputs from two independently trained encoders. The scEEG-Encoder and PPG-Encoder process their respective input modalities and output probability distributions $\mathbf{P}_{\mathrm{scEEG}}$ and $\mathbf{P}_{\mathrm{PPG}}$ over the four sleep stages. These distributions are combined through weighted probability fusion:
\begin{equation}
\mathbf{P}_{\mathrm{score}} = \alpha \cdot \mathbf{P}_{\mathrm{PPG}} + (1-\alpha) \cdot \mathbf{P}_{\mathrm{scEEG}}, 
\label{score_fusion}
\end{equation}
where $\alpha \in [0,1]$ controls modality contribution. Final predictions select the class with maximum fused probability.

\textbf{(b) Cross-attention fusion:} Since the two modalities are time-aligned at the epoch level, cross-attention provides a principled mechanism for modality-conditioned feature reweighting: scEEG queries can select autonomic cues from PPG features, and vice versa, as shown in Fig.~\ref{fig:fusion_architectures}(b). Given extracted features $\mathbf{F}_{\mathrm{scEEG}} \in \mathbb{R}^{T \times d}$ and $\mathbf{F}_{\mathrm{PPG}} \in \mathbb{R}^{T \times d}$ from the frozen encoders (highlighted in Fig.~\ref{fig:eeg}(a) and (b) in red), where $d=256$ is the feature dimension, we apply multi-head cross-attention to allow each modality to attend to the other:
\begin{equation}
\mathbf{F}_{\mathrm{scEEG}}' = \text{CrossAttn}(\mathbf{F}_{\mathrm{scEEG}}, \mathbf{F}_{\mathrm{PPG}}, \mathbf{F}_{\mathrm{PPG}})
\end{equation}
\begin{equation}
\mathbf{F}_{\mathrm{PPG}}' = \text{CrossAttn}(\mathbf{F}_{\mathrm{PPG}}, \mathbf{F}_{\mathrm{scEEG}}, \mathbf{F}_{\mathrm{scEEG}})
\end{equation}
where $\text{CrossAttn}(Q, K, V)$ denotes multi-head cross-attention with query $Q$, key $K$, and value $V$. This bidirectional attention allows scEEG features to incorporate PPG's autonomic signatures for improved light sleep identification, while PPG features leverage scEEG's cortical information for enhanced deep sleep detection. Following standard Transformer design, each cross-attention is followed by a feed-forward network (FFN) with layer normalization and residual connections.

The attended features are processed through two cross-attention fusion blocks, then concatenated and projected to the original dimension:
\begin{equation}
\mathbf{F}_{\mathrm{fused}} = \text{Linear}([\mathbf{F}_{\mathrm{scEEG}}'; \mathbf{F}_{\mathrm{PPG}}'])
\label{crossatten}
\end{equation}
where $[\cdot;\cdot]$ denotes concatenation. The fused features are passed through a classification head to produce final predictions $\mathbf{P}_{\mathrm{fused}} \in \mathbb{R}^{T \times 4}$. During training, we freeze the pre-trained scEEG/PPG encoders from Fig.~\ref{fig:eeg} and only train the cross-attention blocks and classifier, reducing trainable parameters while leveraging the feature extraction capabilities learned from the single-modality models.

\textbf{(c) Mamba-enhanced fusion:} As depicted in Fig.~\ref{fig:fusion_architectures}(c), we extend the cross-attention fusion architecture by incorporating a Mamba-based temporal context module (TCM) to capture sequential dependencies across epochs. While cross-attention enables inter-modal information exchange at each epoch, it does not explicitly model the temporal evolution of sleep stages. We address this limitation by applying selective state space modeling after the cross-attention fusion.

Following Eq.~(\ref{crossatten}), the fused representations $\mathbf{F}_{\mathrm{fused}} \in \mathbb{R}^{T \times d}$ are processed through a bidirectional Mamba TCM. The Mamba block employs selective state space models (SSMs) that dynamically modulate information flow based on input content \cite{gu2024mamba}. For an input sequence $\mathbf{x}_t$ (the $t$-th row of $\mathbf{F}_{\mathrm{fused}}$), the SSM computes:
\begin{equation}
\mathbf{h}_t = \bar{\mathbf{A}}_t \mathbf{h}_{t-1} + \bar{\mathbf{B}}_t \mathbf{x}_t
\end{equation}
\begin{equation}
\mathbf{y}_t = \mathbf{C}_t \mathbf{h}_t
\end{equation}
where $\mathbf{h}_t$ is the hidden state, $\mathbf{y}_t$ is the output at epoch $t$, and the discretized parameters $\bar{\mathbf{A}}_t$, $\bar{\mathbf{B}}_t$, and $\mathbf{C}_t$ are input-dependent, enabling the model to selectively retain or forget information based on sleep-relevant patterns.

To capture both forward and backward temporal context, we employ a bidirectional architecture \cite{lee2024neuronet}. The forward Mamba block $\mathrm{Mamba}_{\rightarrow}$ processes the sequence from $t=1$ to $T$ using Eqs.~(5-6), while the backward block $\mathrm{Mamba}_{\leftarrow}$ processes from $t=T$ to $1$. The bidirectional outputs are then combined:
\begin{equation}
\mathbf{F}_{\mathrm{temporal}} = \mathrm{Merge}(\mathrm{Mamba}_{\rightarrow}(\mathbf{F}_{\mathrm{fused}}), \mathrm{Mamba}_{\leftarrow}(\mathbf{F}_{\mathrm{fused}}))
\end{equation}
where $\mathrm{Merge}$ combines the bidirectional outputs through linear projection. This design allows each epoch's classification to benefit from both preceding and subsequent context, similar to how sleep experts consider adjacent epochs when scoring ambiguous transitions. \rev{Because the module is bidirectional, staging is not strictly causal: an epoch's prediction depends on later epochs within the window, so the effective decision latency is on the order of half a window (${\sim}1.5$\,min for a 3-minute window).}

Finally, the temporal features are passed through a classification head to produce per-epoch predictions $\mathbf{P}_{\mathrm{fused}} \in \mathbb{R}^{T \times 4}$.

\subsection{Evaluation Metrics}
\subsubsection{Performance statistics}
We evaluate model performance using accuracy (Acc) and $\kappa$. Acc measures the proportion of correctly classified epochs. $\kappa$ quantifies agreement between predictions and ground truth while accounting for chance agreement: $\kappa = \frac{p_o - p_e}{1 - p_e}$, where $p_o$ is the proportion of epochs where the model prediction matches the ground truth, and $p_e$ is the expected proportion by random chance, providing a more robust measure for imbalanced datasets \cite{cohen1960coefficient}. \rev{For class-balanced assessment, we additionally report the Matthews correlation coefficient (MCC) and the macro-averaged geometric mean of per-class recall (macro G-mean).} For computational efficiency analysis, we report model parameter counts (Model Size) and per-window inference (Infer.) time. For light sleep evaluation, we report recall and F1-score to assess classification performance on this challenging stage.

\subsubsection{Sleep Measures}
Beyond epoch-level classification metrics, we also compute sleep architecture measures following \cite{attia2025sleepppg}, which characterize the structural organization of sleep through clinically relevant metrics: total sleep time (TST), sleep efficiency (SE), and stage fractions including light sleep (FRLight), deep sleep (FRDeep), and REM (FRREM). These measures are defined as:
\begin{equation}
TST = \sum_{i=1}^{n} Light[i] + \sum_{i=1}^{n} Deep[i] + \sum_{i=1}^{n} REM[i]
\end{equation}
\begin{equation}
SE = \frac{TST}{TST + \sum_{i=1}^{n} Wake[i]} \times 100
\end{equation}
\begin{equation}
FR_{Stage} = \frac{\sum_{i=1}^{n} Stage[i]}{TST} \times 100
\end{equation}

where $n$ is the number of 30-second epochs, and $Stage[i] \in \{0, 1\}$ indicates whether epoch $i$ belongs to the corresponding sleep stage ($Wake$, $Light$, $Deep$, or $REM$). We report mean absolute error (MAE) between predicted and reference values.

\section{Experimental Evaluation}
\subsection{Experimental Setup}

\subsubsection{Window Length Configuration}
To study the minimal window length that achieves acceptable performance for short-window sleep staging, we evaluated six input window lengths: 30s, 1min, 3min, 5min, 10min, and 30min. For each configuration, dedicated scEEG and PPG models were trained, as illustrated in Fig.~\ref{fig:eeg}(a) and (b), respectively. The scEEG model extends AttnSleep \cite{eldele2021attention} with a temporal transformer for cross-epoch modeling. The PPG model follows the architecture proposed in \cite{wang2025improving}, with the encoder depth adjusted according to window length such that shorter windows use fewer convolutional layers. Both models were trained on the MESA dataset.

\begin{figure*}[htbp]
    \centering
    \includegraphics[width=0.9\textwidth]{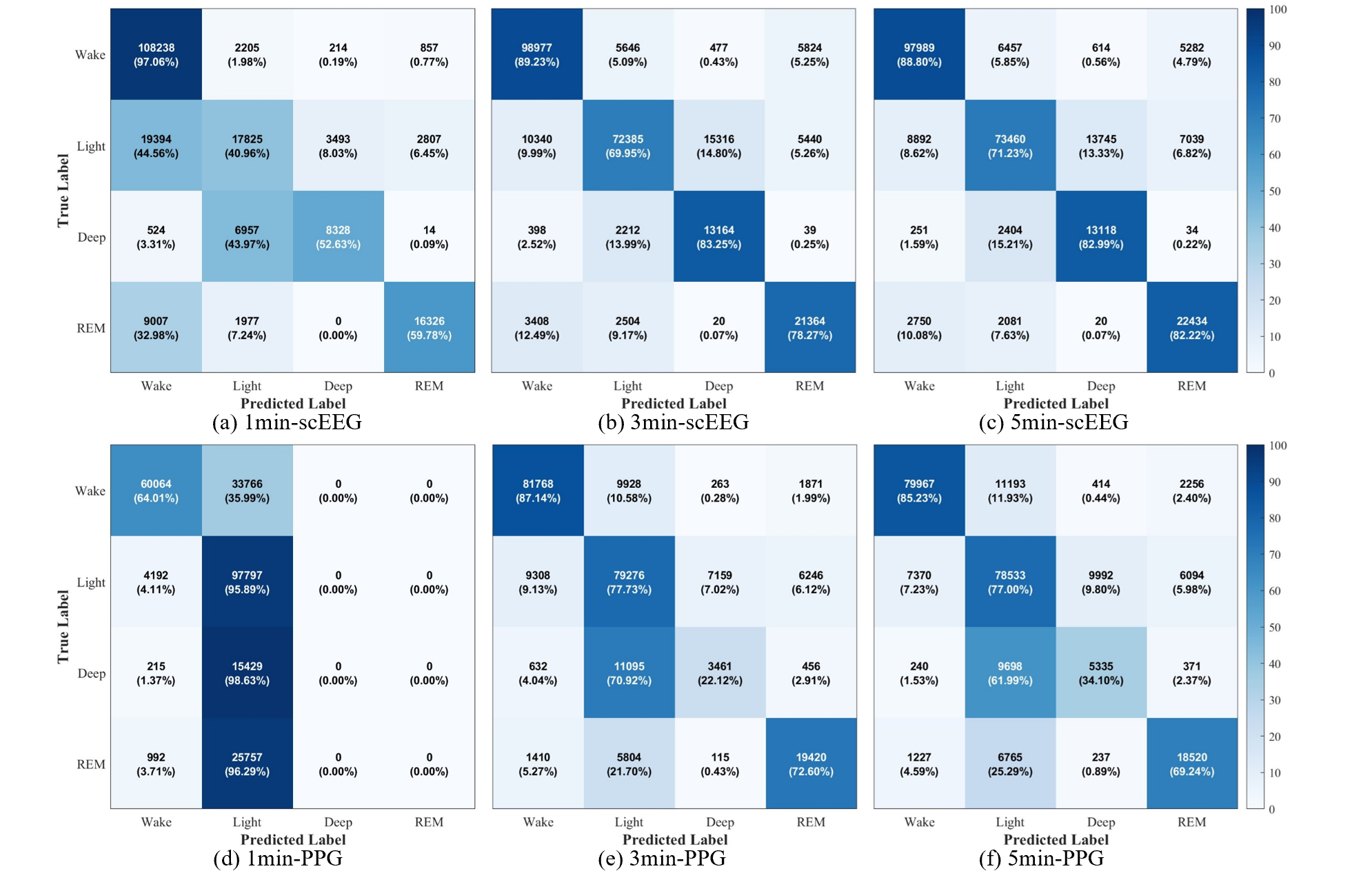}
    \caption{Confusion matrices for sleep stage classification across different window lengths using scEEG and PPG.}
    \label{fig:window}
\end{figure*}

\subsubsection{Training Protocol} Both modality-specific models were trained under a sequence-labeling objective with \rev{a class-weighted cross-entropy loss}. The train/validation/test split followed the SleepPPG-Net protocol \cite{kotzen2022sleepppg}, with 204 subjects reserved for testing. All experiments were implemented in PyTorch 2.0 and executed on NVIDIA RTX 4090 GPUs.
For fusion experiments, we investigated three strategies with different training requirements:
(a) Score-level fusion: The fusion weight was selected via grid search over $\{0, 0.1, 0.2, ..., 1.0\}$ on a held-out validation set, where the fused prediction is defined in Eq.~(\ref{score_fusion}). (b) Cross-attention fusion: The scEEG and PPG encoders (Fig.~\ref{fig:eeg}(a) and (b)) were frozen, and only the cross-attention layers and classifier (\rev{3.32M} trainable parameters) were trained for \rev{20 epochs with learning rate $1 \times 10^{-4}$}. (c) Mamba-enhanced fusion: Building upon cross-attention fusion, the bidirectional Mamba TCM adds \rev{1.88M} trainable parameters. Training followed the same protocol as cross-attention fusion.

\subsubsection{Implementation Details}
\rev{\textbf{Preprocessing.} scEEG signals are band-pass filtered at
$0.3$--$35$\,Hz with a zero-phase FIR filter (MNE \texttt{firwin} design) and
resampled to $100$\,Hz by polyphase resampling ($3000$ samples per $30$\,s epoch),
followed by per-window $z$-score normalization. PPG signals are low-pass filtered
by an $8$th-order Chebyshev-II filter at $8$\,Hz, resampled to $34.13$\,Hz
($1024$ samples per epoch), clipped at $\pm 3\sigma$, and recording-level
$z$-score normalized.

\textbf{Architecture.} Dropout is $0.5$ in the scEEG MRCNN front-end and $0.1$ in
its temporal transformer; $0.1$--$0.2$ in the PPG encoder; and $0.1$ throughout
the cross-attention and Mamba fusion blocks. The Mamba TCM uses
$d_{\text{state}}\!=\!16$, $d_{\text{conv}}\!=\!4$, expansion factor $2$, and
$2$ layers ($2$ cross-attention fusion blocks).

\textbf{Training.} The unimodal scEEG and PPG encoders are each trained with AdamW
(weight decay $10^{-5}$, batch size $32$, $15$ epochs, early-stopping patience $5$,
gradient clipping $1.0$) under a class-weighted cross-entropy loss (inverse-frequency
weights); the scEEG encoder uses learning rate $10^{-3}$ with a OneCycle schedule
and the PPG encoder $3\times10^{-4}$ with cosine annealing. Fusion models are trained
on frozen encoders with AdamW (learning rate $10^{-4}$, weight decay $10^{-4}$,
cosine annealing, batch size $32$, $20$ epochs, early-stopping patience $10$) and a
square-root inverse-frequency class-weighted cross-entropy loss. Cross-dataset
fine-tuning uses a reduced learning rate of $10^{-5}$ with early stopping on the
target validation set. All models use random seed $42$; the main MESA fusion
comparison is additionally repeated over $3$ seeds. Model selection uses the best
validation $\kappa$.}

\subsubsection{Cross-Dataset Validation Protocol}
To assess generalization capability, we evaluated our models on CFS and ABC datasets using two paradigms: (1) direct transfer, where MESA-trained models were directly applied without adaptation, and (2) fine-tuning, where MESA-trained models were further optimized on target training data with a reduced learning rate of $1 \times 10^{-5}$ and early stopping based on target validation performance. For both datasets, PPG signals were resampled to 34.13 Hz (1,024 samples per epoch) and scEEG signals were processed at 100 Hz (3,000 samples per epoch) to match MESA preprocessing. For fusion models, we adopted the fine-tuning paradigm, as the cross-attention and Mamba modules require adaptation to capture dataset-specific inter-modal relationships. \rev{All splits are performed at the subject level, so no subject appears in more than one split. The target-domain subjects ($719$ for CFS, $49$ for ABC) are partitioned into train/validation/test sets with a fixed random seed, using an $80/10/10$ ratio for the fusion models and $64/16/20$ for the single-modality models.}

\subsubsection{Baselines}
We compare our fusion approaches against SOTA unimodal methods:

\begin{table}[t]
\centering
\caption{scEEG model performance across window lengths on MESA test set (M=204).}
\label{tab:eeg_short_window_performance}
\begin{threeparttable}
\begin{tabular}{lcccc}
\toprule
\textbf{Window} & \textbf{$\kappa$} & \textbf{Acc} & \textbf{Model Size} & \textbf{Infer. (ms)} \\
\midrule
30 s  & 0.665 & 0.772 & 0.52M & 3.84 $\pm$ 0.23 \\
1 min & 0.674 & 0.783 & 2.71M & 3.92 $\pm$ 0.24 \\
3 min & 0.697 & 0.799 & 2.71M & 3.97 $\pm$ 0.18 \\
5 min & 0.721 & 0.815 & 2.71M & 3.98 $\pm$ 0.26 \\
10 min & 0.688 & 0.787 & 2.71M & 3.94 $\pm$ 0.27 \\
30 min & 0.735 & 0.821 & 2.71M & 3.94 $\pm$ 0.32 \\
\bottomrule
\end{tabular}
\begin{tablenotes}[flushleft]
\footnotesize
\item[$\dagger$] Model size is constant for windows $\geq$ 1 min; the 30\,s model uses a single-epoch setting without a temporal transformer.
\end{tablenotes}
\end{threeparttable}
\end{table}

\begin{table}[t]
\centering
\caption{PPG model performance across window lengths on the MESA test set (M=204).}
\label{tab:short_window_performance}
\begin{threeparttable}
\begin{tabular}{lcccc}
\toprule
\textbf{Window} & \textbf{$\kappa$} & \textbf{Acc} & \textbf{Model Size} & \textbf{Infer.(ms)} \\
\midrule
30 s  & 0.460 & 0.647 & 8.28M  & 8.44 $\pm$ 0.56 \\
1 min & 0.538 & 0.701 & 8.29M  & 9.10 $\pm$ 0.68 \\
3 min & 0.643 & 0.772 & 10.26M & 9.39 $\pm$ 0.39 \\
5 min & 0.633 & 0.758 & 10.26M & 9.41 $\pm$ 0.55 \\
10 min & 0.656 & 0.777 & 12.23M & 10.28 $\pm$ 0.59 \\
30 min & 0.696 & 0.804 & 14.20M & 11.05 $\pm$ 0.60 \\
\bottomrule
\end{tabular}

\begin{tablenotes}[flushleft]
\footnotesize
\item[$\dagger$] Encoder depth increases with window length: 4 layers (30\,s), 5 layers (1\,min), 6 layers (3--5\,min), 7 layers (10\,min), and 8 layers (30\,min).
\end{tablenotes}
\end{threeparttable}
\end{table}

\textbf{PPG-based:} (1) SleepPPGNet~\cite{kotzen2022sleepppg}, originally designed for full-night PPG recordings, processes signals through residual convolutional blocks and temporal convolution layers for 4-class sleep staging. We adapted this architecture to short-window inputs by reducing the input sequence length while preserving the core network structure. (2) Dual-stream PPG employs cross-attention fusion between raw and augmented PPG streams to extract complementary features~\cite{wang2025improving}.

\begin{table*}[t]
\centering
\caption{Comparison of fusion models and SOTA methods under the 3-minute window on the MESA dataset.}
\label{tab:performance_comparison}
\setlength{\tabcolsep}{6pt}
\renewcommand{\arraystretch}{1.2}
\begin{tabular}{llcccc}
\toprule
\textbf{Strategy} & \textbf{Method} 
& $\boldsymbol{\kappa}$ 
& \textbf{Acc} 
& \textbf{Model Size} 
& \textbf{Infer. (ms)} \\
\midrule
\multirow{2}{*}{PPG}
 & SleepPPGNet \cite{kotzen2022sleepppg} 
 & 0.479 & 0.659 & 2.29M & 8.10 $\pm$ 0.35 \\
 & Dual-stream PPG \cite{wang2025improving} 
 & 0.643 & 0.772 & 10.26M & 9.39 $\pm$ 0.39 \\
\midrule
\multirow{2}{*}{scEEG}
 & AttnSleep \cite{eldele2021attention} 
 & 0.697 & 0.799 & 2.71M & 3.97 $\pm$ 0.18 \\
 & CBraMod \cite{wang2024cbramod} 
 & 0.762 & 0.769 & 12.14M & 6.18 $\pm$ 0.79 \\
\midrule
\multirow{3}{*}{Fusion}
 & Score-level (ours)  
 & 0.747 & 0.835 & 12.9M  & 14.40 $\pm$ 1.07 \\
 & Cross-attention (ours)
 & 0.767 & 0.844 & 16.29M & 15.69 $\pm$ 1.27 \\
 & \textbf{Mamba-enhanced (ours)} 
 & \textbf{0.798} & \textbf{0.869} & 18.17M & 19.98 $\pm$ 2.37 \\
\bottomrule
\end{tabular}
\end{table*}

\textbf{scEEG-based:} (1) AttnSleep uses multi-resolution CNNs to capture both fine-grained and coarse temporal patterns~\cite{eldele2021attention}, followed by transformer-based temporal context encoding with multi-head self-attention. (2) CBraMod is a foundation model pre-trained on large-scale EEG data using self-supervised learning~\cite{wang2024cbramod}. We used the single-channel configuration and fine-tuned the model on MESA data for 4-class sleep staging using the same data preprocessing as our scEEG model.

\subsection{Results}

\textbf{scEEG Model Performance.}
The scEEG model shows improved performance as input window length increases, with $\kappa$ ranging from 0.665 (30\,s) to 0.735 (30\,min), as summarized in Table~\ref{tab:eeg_short_window_performance}. We present detailed confusion matrices for 1, 3, and 5-minute windows in Fig.~\ref{fig:window}(a-c), as these represent the practical range for short-window applications while capturing the key performance transitions. As illustrated in the confusion matrices, longer windows improve overall classification consistency.

\textbf{PPG Model Performance.}
In contrast, the PPG model shows a stronger dependence on input window length. As summarized in Table~\ref{tab:short_window_performance}, ultra-short windows (30\,s and 1\,min) result in limited discriminative capability, reflecting insufficient temporal context to capture slow-varying cardiovascular dynamics. Performance improves substantially at 3-minute windows ($\kappa=0.643$), while longer windows yield only modest additional gains at the cost of increased model complexity. Fig.~\ref{fig:window}(d-f) presents detailed confusion matrices for 1, 3, and 5-minute windows.

\textbf{Window Length Selection for Fusion.}
Based on these observations, we adopt a 3-minute window for subsequent scEEG-PPG fusion experiments. For scEEG, this window preserves sufficient temporal resolution for tracking sleep stage transitions while maintaining competitive classification performance. For PPG, windows shorter than 3 minutes lead to pronounced performance degradation, whereas a 3-minute window represents a reasonable duration that achieves stable and reliable classification results. Critically, a 3-minute window allows the system to provide a staging decision within minutes of sleep onset rather than requiring hours of data accumulation. This shared window also enables synchronized multimodal fusion with a balanced trade-off between classification performance and computational efficiency.

\textbf{Fusion Method Comparison.}
Table~\ref{tab:performance_comparison} compares our fusion approaches against SOTA unimodal methods on the MESA dataset using 3-minute windows. PPG-based methods achieve $\kappa$ ranging from 0.479 (SleepPPGNet) to 0.643 (Dual-stream PPG), while scEEG-based methods reach 0.697 (AttnSleep) to 0.762 (CBraMod). Although CBraMod achieves the highest unimodal $\kappa$, its model size (12.14M parameters) is over five times larger than AttnSleep (2.71M), and its accuracy (76.9\%) is lower than AttnSleep (79.9\%). Given the trade-off between lightweight requirements and performance, we adopted AttnSleep as the scEEG encoder for our fusion strategies.

\begin{figure}[t]
    \centering
    \includegraphics[width=\columnwidth]{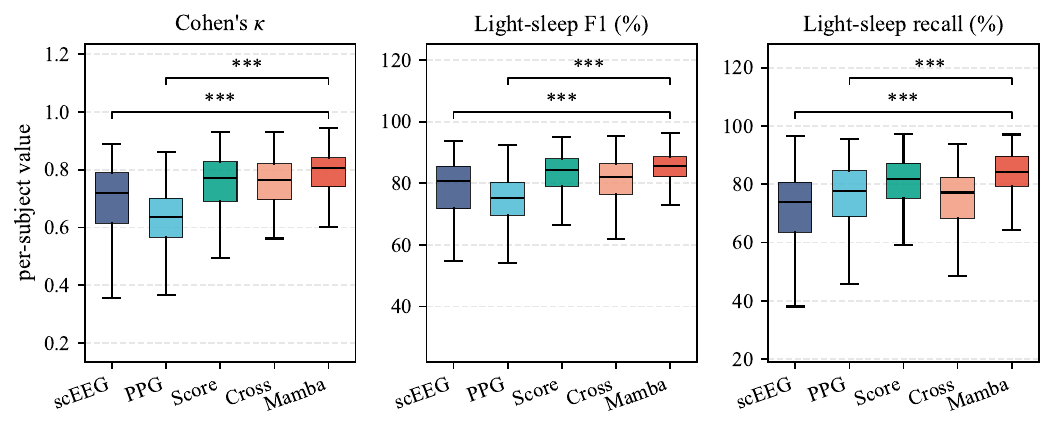}
    \caption{\rev{Per-subject Cohen's $\kappa$, light-sleep F1, and recall on MESA ($M\!=\!204$); $^{***}$: $p<0.001$ (Wilcoxon) for Mamba-enhanced fusion vs.\ each single modality.}}
    \label{fig:stats_boxplot}
\end{figure}

\begin{figure}[htbp]
    \centering
    \includegraphics[width=0.85\columnwidth]{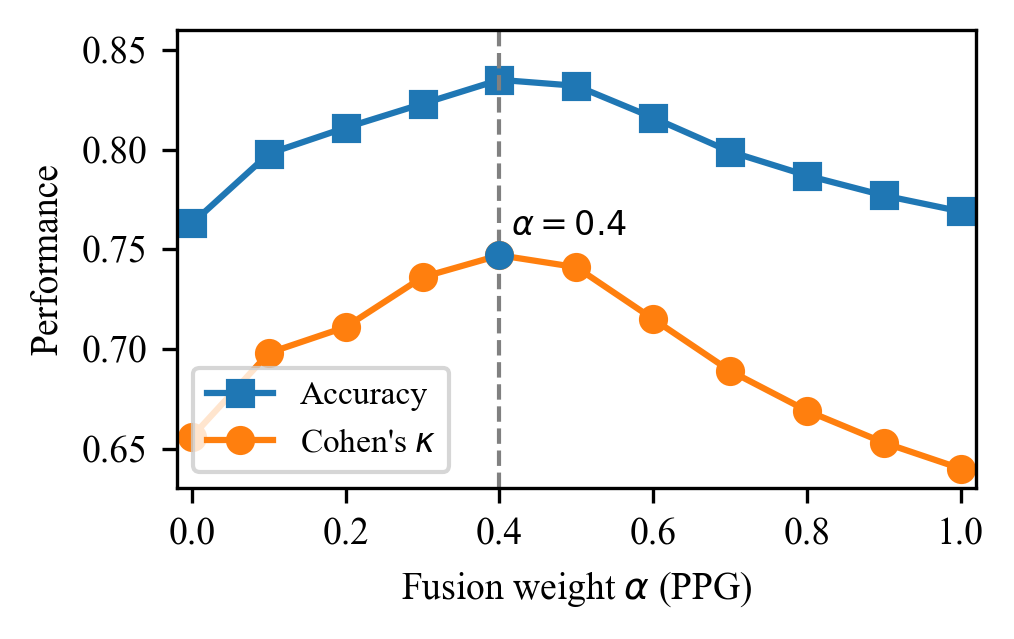}
    \caption{Effect of fusion weight $\alpha$ on performance under the 3-minute window.}
    \label{fig:fusion_weight}
\end{figure}

Building upon the unimodal baselines, we investigate three fusion strategies. Score-level fusion achieves $\kappa=0.747$, surpassing most unimodal methods. As shown in Fig.~\ref{fig:fusion_weight}, the fusion weight $\alpha=0.4$ yields the best performance, indicating that scEEG contributes more discriminative information than PPG for sleep staging. Cross-attention fusion further improves to $\kappa=0.767$ by enabling the model to learn which features from one modality are most relevant given the other. Mamba-enhanced fusion achieves the best overall performance ($\kappa=0.798$, $Acc=86.9\%$). \rev{We adopt Bi-Mamba for the temporal context module (parameter-matched ablation below); the gains are driven primarily by the multimodal fusion rather than the specific temporal architecture.}
\rev{Subject-level paired tests confirm the fusion significantly outperforms each
single modality. Over the 204 MESA test subjects (subject-bootstrap $95\%$ CIs,
$B=10{,}000$; Holm-corrected Wilcoxon), Mamba-enhanced fusion improves per-subject
$\kappa$ by $+0.114$ over scEEG ($[0.097, 0.133]$; $p<0.001$; $d_z=0.87$) and by
$+0.161$ over PPG ($[0.148, 0.175]$; $p<0.001$; $d_z=1.65$), and attains the highest
MCC ($0.798$) and macro G-mean ($0.814$); per-method distributions are shown in
Fig.~\ref{fig:stats_boxplot}. Retraining over three seeds yields
$\kappa=0.7984\pm0.0005$.}

\textbf{Temporal Module Ablation.}
\rev{To test whether the temporal context module (TCM) must be a selective
state-space (Mamba) model, we ran a parameter-matched ablation replacing the
Bi-Mamba TCM with a Transformer, a bidirectional GRU, or a bidirectional LSTM,
holding the encoders, fusion, classifier, and training recipe fixed and matching
trainable parameters ($\approx\!5.2$\,M; Table~\ref{tab:tcm_ablation}). The four
modules perform comparably (per-subject $\kappa$ $0.781$--$0.787$, overlapping
$95\%$ CIs), so staging performance is not specific to the Mamba formulation;
the state-space and attention modules marginally exceed the gated RNNs. We adopt
Bi-Mamba as it attains the highest per-subject $\kappa$ while uniquely offering
linear-time sequence modelling.}

\begin{table}[t]\centering
\caption{\rev{Parameter-matched temporal-module ablation on the MESA test set, where only the TCM differs with trainable parameters matched; per-subject $\kappa$ is mean [95\% CI].}}
\label{tab:tcm_ablation}
\setlength{\tabcolsep}{4pt}
\begin{tabular}{lcc}
\toprule
\textbf{Temporal module} & \textbf{Train. params} & $\boldsymbol{\kappa}$ \textbf{(95\% CI)} \\
\midrule
Transformer      & 5.21\,M & 0.786 [0.775, 0.798] \\
Bi-GRU           & 5.20\,M & 0.783 [0.774, 0.793] \\
Bi-LSTM          & 5.22\,M & 0.781 [0.772, 0.791] \\
\rev{Bi-Mamba}  & 5.21\,M & 0.787 [0.777, 0.798] \\
\bottomrule
\end{tabular}
\end{table}

\begin{figure*}[htbp]
    \centering
    \includegraphics[width=\textwidth]{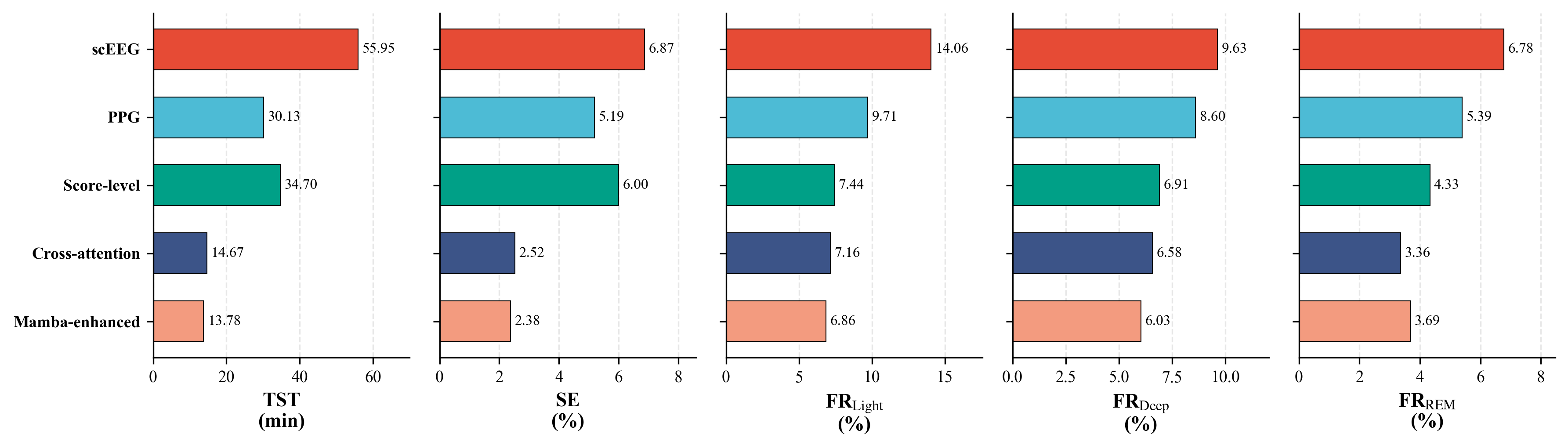}
    \caption{Sleep architecture estimation errors (MAE) on MESA test set.}
    \label{fig:sleep_measures}
\end{figure*}

\textbf{Sleep Architecture Estimation.} 
Accurate sleep architecture estimation is essential for clinical sleep assessment and long-term health monitoring. As shown in Fig.~\ref{fig:sleep_measures}, Mamba-enhanced fusion achieves the lowest estimation errors across most metrics, reducing TST MAE from 55.95\,min (scEEG) and 30.13\,min (PPG) to 13.78\,min \rev{(95\% CI $[11.9,15.9]$)}, and SE MAE from 6.87\% (scEEG) and 5.19\% (PPG) to 2.38\% \rev{(95\% CI $[2.1,2.7]$)}. These substantial reductions in estimation error demonstrate that multimodal fusion could provide more reliable sleep architecture metrics for home-based lightweight sleep monitoring.

\begin{figure}[t]
    \centering
    \includegraphics[width=\columnwidth]{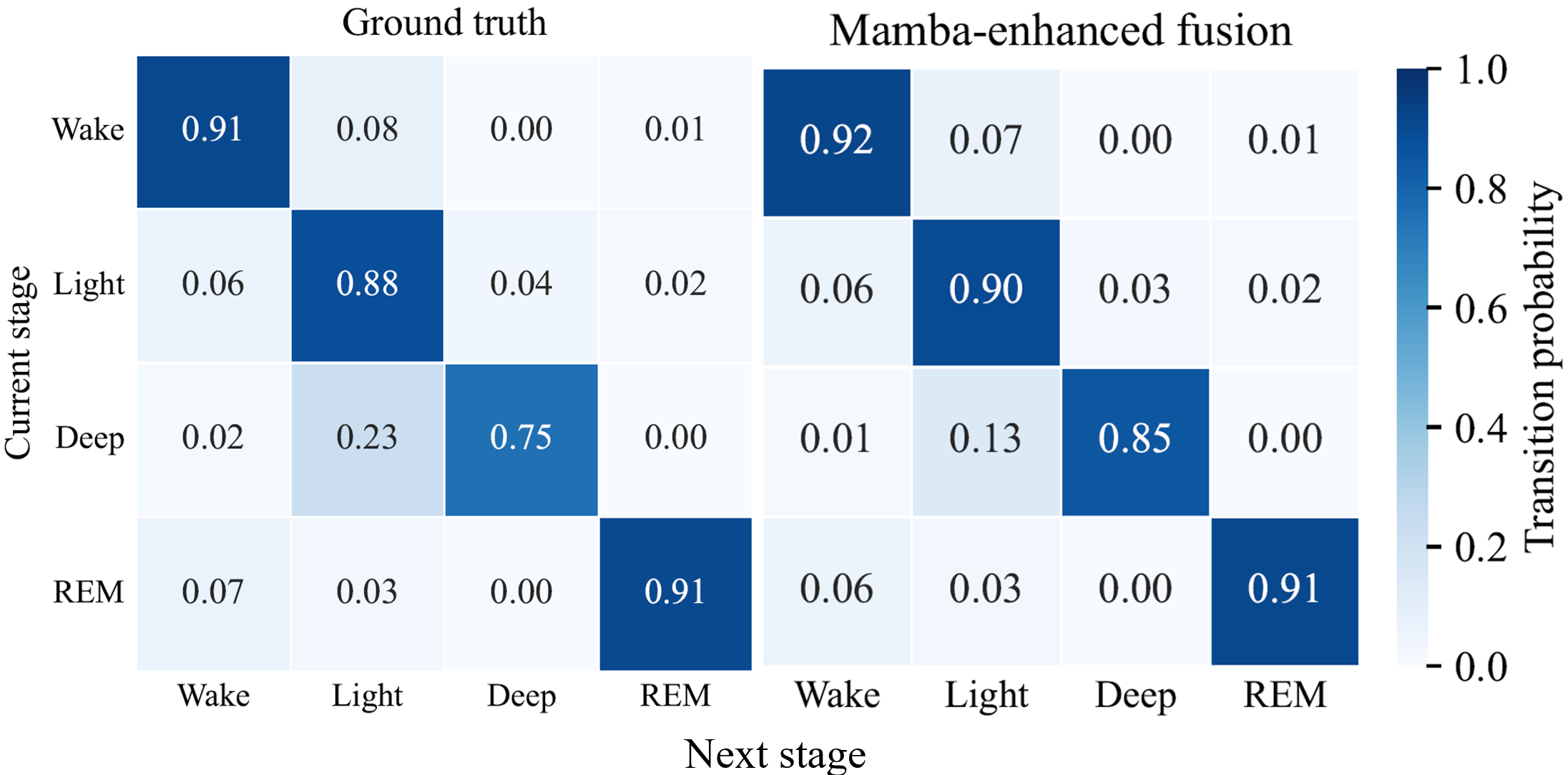}
    \caption{\rev{Sleep-stage transition-probability matrices: ground truth (left) vs. Mamba-enhanced fusion (right) on MESA.}}
    \label{fig:transition}
\end{figure}

\rev{\textbf{Sleep-Stage Dynamics.} A clinically useful model should also reproduce
the temporal structure of sleep. Fig.~\ref{fig:transition} compares the
Wake/Light/Deep/REM transition-probability matrices of the ground-truth and
predicted hypnograms on the MESA test set. The predicted dynamics closely match
ground truth (mean absolute difference $0.015$ over the $16$ transitions):
self-persistence and implausible transitions ($\text{Wake}\!\to\!\text{Deep}$,
$\text{Deep}\!\to\!\text{REM}\approx0$) are faithfully reproduced, with only mild
deep-sleep over-persistence ($\text{Deep}\!\to\!\text{Deep}$ $0.85$ vs.\ $0.75$).}

\textbf{\rev{Subgroup Analysis.}}
\rev{We break the adopted fusion model down on the $204$-subject test set by OSA
severity, age, sex, and race/ethnicity (Table~\ref{tab:subgroup}). Performance is
consistent throughout (per-subject $\kappa$ $0.765$--$0.828$). As expected
clinically, $\kappa$ declines modestly with greater OSA severity ($0.828$ normal to
$0.775$ severe) and age ($0.799$ at $65$--$74$\,yr to $0.776$ at $\geq\!75$\,yr),
is essentially identical between sexes, and similar across race/ethnicity (the
lowest, $0.765$, being the smallest subgroup, $n\!=\!21$), supporting reliability
across heterogeneous populations.}

\begin{table}[t]\centering
\caption{\rev{Subgroup analysis of the Mamba-enhanced fusion on the MESA test set
($n\!=\!204$): per-subject $\kappa$ (mean [95\% CI]) and accuracy.}}
\label{tab:subgroup}
\setlength{\tabcolsep}{4pt}
\renewcommand{\arraystretch}{1.1}
\begin{tabular}{llccc}
\toprule
\textbf{Factor} & \textbf{Subgroup} & \textbf{n} & $\boldsymbol{\kappa}$ \textbf{(95\% CI)} & \textbf{Acc (\%)} \\
\midrule
\multirow{4}{*}{OSA (AHI)}
 & Normal ($<5$)         & 17  & 0.828 [0.795, 0.853] & 89.1 \\
 & Mild ($5$--$15$)      & 64  & 0.786 [0.765, 0.806] & 86.6 \\
 & Moderate ($15$--$30$) & 66  & 0.788 [0.769, 0.807] & 86.7 \\
 & Severe ($\geq30$)     & 57  & 0.775 [0.757, 0.791] & 86.5 \\
\midrule
\multirow{3}{*}{Age (yr)}
 & $<65$      & 67  & 0.787 [0.769, 0.805] & 86.6 \\
 & $65$--$74$ & 67  & 0.799 [0.781, 0.817] & 87.7 \\
 & $\geq75$   & 70  & 0.776 [0.757, 0.794] & 86.2 \\
\midrule
\multirow{2}{*}{Sex}
 & Male   & 99  & 0.789 [0.774, 0.804] & 87.3 \\
 & Female & 105 & 0.785 [0.770, 0.799] & 86.4 \\
\midrule
\multirow{4}{*}{Race}
 & White    & 64 & 0.795 [0.777, 0.811] & 87.3 \\
 & Black    & 58 & 0.788 [0.768, 0.808] & 87.0 \\
 & Hispanic & 61 & 0.786 [0.766, 0.805] & 86.7 \\
 & Asian    & 21 & 0.765 [0.728, 0.798] & 85.6 \\
\bottomrule
\end{tabular}
\end{table}

\begin{figure}[t]
    \centering
    \includegraphics[width=\columnwidth]{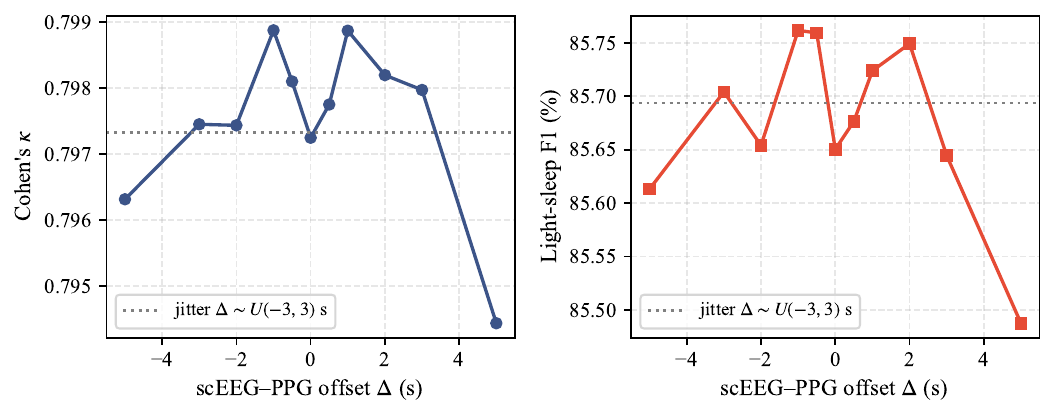}
    \caption{\rev{Synchronization sensitivity: $\kappa$ and light-sleep F1 of the Mamba-enhanced fusion versus an imposed scEEG-PPG temporal offset $\Delta$ (MESA).}}
    \label{fig:sync}
\end{figure}

\textbf{\rev{Synchronization Sensitivity.}}
\rev{Because consumer wearables may not share a common clock across sensors, we
assess robustness to misalignment by re-evaluating the adopted model (without
retraining) while shifting the PPG stream relative to scEEG by an offset $\Delta$.
Performance is essentially flat across $\Delta\in\{\pm0.5,\pm1,\pm2,\pm3\}$\,s---the
realistic clock-drift range---with $\kappa$ within $[0.797,0.799]$, indistinguishable
from the synchronized baseline ($0.797$); a large $\pm5$\,s offset costs only
$\approx\!0.003$, and per-window random jitter ($\Delta\sim\mathcal{U}(-3,3)$\,s)
leaves $\kappa$ at $0.797$ (Fig.~\ref{fig:sync}). This tolerance arises because
cross-attention aligns the modalities at the $3$-minute window scale rather than
sample-by-sample. Missing data and motion artefacts under fully asynchronous
acquisition remain deployment considerations noted in the Limitations.}

\textbf{Cross-dataset Evaluation.}
To evaluate generalization capability, we apply MESA-pretrained models to CFS and ABC datasets under two paradigms: without fine-tuning and with fine-tuning.

\begin{table}[htbp]
\caption{Single-modality generalization performance across datasets.}
\centering
\begin{tabular}{lcccc}
\toprule
\multirow{2}{*}{Strategy} 
& \multicolumn{2}{c}{CFS} 
& \multicolumn{2}{c}{ABC} \\
\cmidrule(r){2-3} \cmidrule(r){4-5}
& $\kappa$ & Acc 
& $\kappa$ & Acc \\
\midrule
PPG (w/o fine-tuning) & 0.090  & 0.432 & 0.368 & 0.66 \\
scEEG (w/o fine-tuning)      & 0.638  & 0.741 & 0.334 & 0.535 \\
\midrule
PPG (fine-tuning)         & 0.583 & 0.710 & 0.494 & 0.691 \\
scEEG (fine-tuning)      & 0.716  & 0.799 & 0.543 & 0.681 \\
\bottomrule
\end{tabular}
\label{tab:cross_dataset-1}
\end{table}

Table~\ref{tab:cross_dataset-1} presents single-modality generalization results. Without fine-tuning, scEEG demonstrates strong transfer on CFS ($\kappa=0.638$) but degrades substantially on ABC ($\kappa=0.334$), while PPG shows the opposite pattern with poor CFS transfer ($\kappa=0.090$) but relatively better ABC transfer ($\kappa=0.368$). This asymmetry indicates that each modality exhibits dataset-specific sensitivity, with neither consistently generalizing across all populations. After fine-tuning, both modalities improve considerably: scEEG achieves $\kappa=0.716$ on CFS and $\kappa=0.543$ on ABC, while PPG reaches $\kappa=0.583$ on CFS and $\kappa=0.494$ on ABC. These results suggest that fine-tuning remains necessary for adapting single-modality models to new populations.

\begin{table}[htbp]
\centering
\caption {\rev{Cross-dataset generalization of fusion methods. Rows are fine-tuned unless marked zero-shot (zs; direct transfer with no target-domain training.)}}
\label{tab:fusion_cross_dataset}
\setlength{\tabcolsep}{3pt}
\renewcommand{\arraystretch}{1.25}
\begin{tabular}{llcccc}
\toprule
\textbf{Dataset} & \textbf{Fusion Method}
& $\boldsymbol{\kappa}$ 
& \textbf{Acc} 
& \textbf{Model Size} 
& \textbf{Infer. (ms)} \\
\midrule

\multirow{5}{*}{CFS}
 & Score-level        & 0.660 & 0.772 & 12.9M  & 14.62 $\pm$ 1.25 \\
 & Cross-attention    & 0.781 & 0.849 & 16.29M & 16.05 $\pm$ 1.14 \\
 & Mamba-enhanced     & 0.769 & 0.839 & 18.17M & 19.20 $\pm$ 1.47 \\
 & \rev{Cross-attention (zs)} & \rev{0.235} & \rev{0.506} & 16.29M & 16.05 $\pm$ 1.14 \\
 & \rev{Mamba-enhanced (zs)}  & \rev{0.152} & \rev{0.462} & 18.17M & 19.20 $\pm$ 1.47 \\
\midrule

\multirow{5}{*}{ABC}
 & Score-level        & 0.524 & 0.684 & 12.9M  & 13.38 $\pm$ 0.56 \\
 & Cross-attention    & 0.643 & 0.773 & 16.29M & 16.13 $\pm$ 1.17 \\
 & Mamba-enhanced     & 0.665 & 0.764 & 18.17M & 19.71 $\pm$ 1.66 \\
 & \rev{Cross-attention (zs)} & \rev{0.552} & \rev{0.710} & 16.29M & 16.13 $\pm$ 1.17 \\
 & \rev{Mamba-enhanced (zs)}  & \rev{0.544} & \rev{0.712} & 18.17M & 19.71 $\pm$ 1.66 \\
\bottomrule
\end{tabular}
\end{table}

\begin{figure*}[t]
    \centering
    \includegraphics[width=0.75\textwidth]{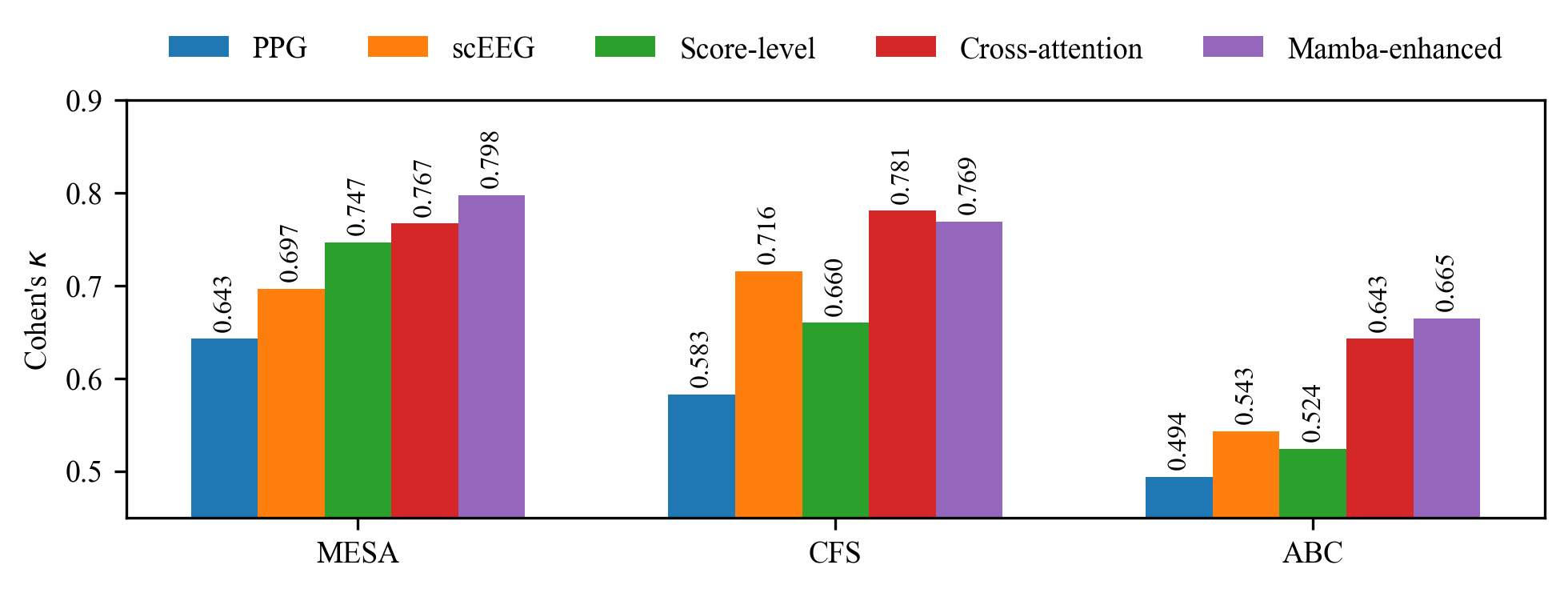}
    \caption{Comparison of $\kappa$ across single-modality and fusion-based methods on the MESA, CFS, and ABC datasets.}
    \label{fig:kappa_comparison}
\end{figure*}

Table~\ref{tab:fusion_cross_dataset} summarizes fusion model performance after fine-tuning. On CFS, cross-attention fusion achieves the highest performance ($\kappa=0.781$, $Acc=84.9\%$), outperforming Mamba-enhanced fusion ($\kappa=0.769$), possibly because the relatively small fine-tuning dataset favors the simpler cross-attention architecture over the Mamba-enhanced model with more trainable parameters. As shown in Fig.~\ref{fig:kappa_comparison}, all fusion methods substantially exceed the best single-modality result ($\kappa=0.716$), confirming the benefit of multimodal integration. \rev{The adopted Mamba-enhanced fusion carries subject-bootstrap $95\%$ CIs of $[0.738,0.798]$ on CFS ($n=73$ test subjects) and $[0.535,0.772]$ on ABC ($n=6$); the markedly wider ABC interval reflects its small test set and is consistent with the modest, cohort-dependent cross-dataset gains.}

\begin{figure}[t]
    \centering
    \includegraphics[width=0.75\columnwidth]{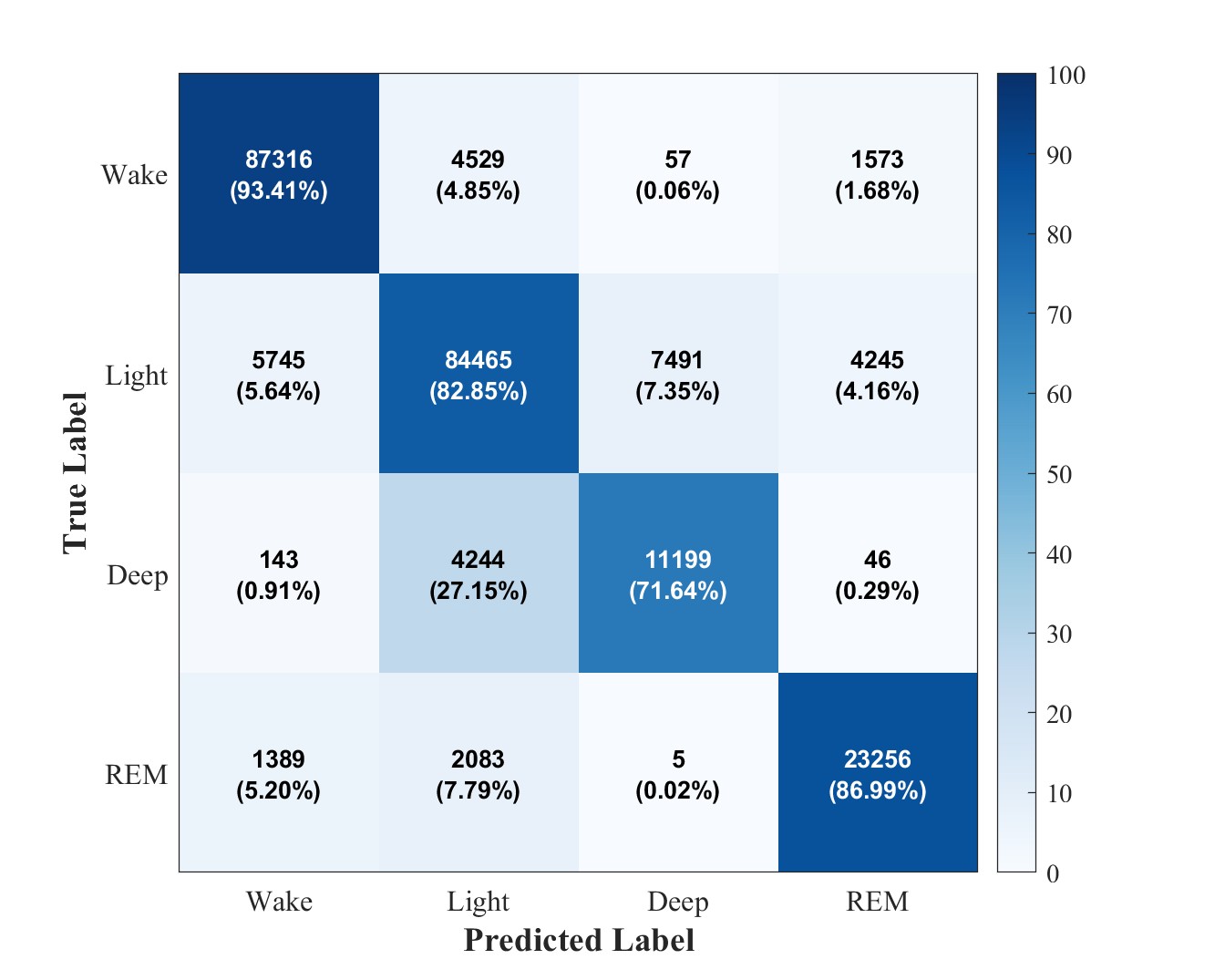}
    \caption{Confusion matrix of the Mamba-enhanced fusion model on the MESA dataset.}
    \label{fig:confusion_mamba_mesa}
\end{figure}

On ABC, Mamba-enhanced fusion demonstrates the best generalization ($\kappa=0.665$, $Acc=76.4\%$), followed by cross-attention fusion ($\kappa=0.643$). The relatively lower performance compared to CFS reflects greater domain shift, as ABC has a narrower age range (26 - 64 years) and limited sample size (49 subjects), making adaptation more challenging, as shown in Fig.~\ref{fig:demographics}. Nevertheless, fusion methods consistently outperform single-modality methods as illustrated in Fig.~\ref{fig:kappa_comparison}, demonstrating that learned fusion strategies provide robust cross-dataset transfer.

\rev{To separate pure transfer from adaptation, Table~\ref{tab:fusion_cross_dataset}
also reports zero-shot (no fine-tuning) fusion; model size and latency are unchanged
since only the weights differ. Direct transfer is dataset-dependent: on ABC it retains
most of the fine-tuned performance ($\kappa=0.55$ vs.\ $0.64$ for cross-attention),
whereas on CFS it collapses ($\kappa=0.24$ and $0.15$) because the CFS PPG distribution
is far from MESA---single-modality PPG direct transfer is itself only $\kappa=0.09$
(Table~\ref{tab:cross_dataset-1}). Thus the MESA-calibrated cross-modal weighting needs
light target-domain adaptation to transfer; being training-free, the zero-shot results
are also free of any target-side subject- or family-level leakage.}

\section{Discussion}
\label{sec:discussion}
\subsection{Physiological Interpretation of Modality Complementarity} 
A key challenge in scEEG-based sleep staging is the unreliable classification of light sleep, as evidenced by the confusion matrices in Fig.~\ref{fig:window}(b). The scEEG model achieves light sleep recall of only 69.95\% and F1-score of 77.76\%, substantially lower than other stages. This disparity reflects the physiological basis that light sleep exhibits subtle, low-amplitude mixed-frequency EEG activity that overlaps with wakefulness and REM patterns~\cite{patel2024physiology}, making it difficult to distinguish from a single EEG channel.

The PPG model exhibits a complementary performance pattern. As shown in Fig.~\ref{fig:window}(e), PPG achieves higher light sleep recall (77.73\%) than scEEG (69.95\%). This advantage aligns with physiological principles: light sleep (N1+N2) is accompanied by a gradual shift toward parasympathetic dominance~\cite{kim2022autonomic}, producing characteristic autonomic changes that PPG can capture through pulse morphology and heart rate variability~\cite{park2022photoplethysmogram}.

Our fusion results confirm this complementarity: the Mamba-enhanced fusion raises light sleep recall to 82.85\% and F1-score to 85.63\% (Fig.~\ref{fig:confusion_mamba_mesa}), showing that PPG effectively compensates for scEEG's light-sleep limitation.

\begin{figure}[t]
    \centering
    \includegraphics[width=\columnwidth]{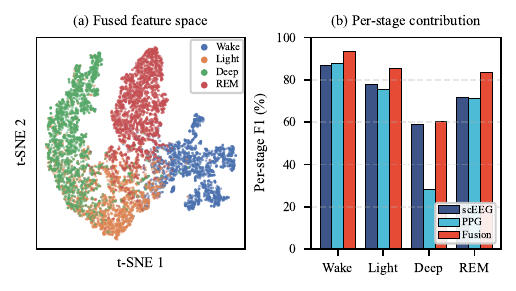}
    \caption{\rev{Fusion interpretability (MESA): (a) t-SNE of the fused features
by stage; (b) per-stage F1.}}
    \label{fig:modality}
\end{figure}

  \rev{The fusion learns a stage-discriminative representation: a t-SNE projection of the
  256-d fused features (Fig.~\ref{fig:modality}(a)) separates wake, deep, and REM into
  distinct clusters, with the transitional light sleep dispersed between them, consistent
  with its known difficulty and with the confusion matrices.
  Fig.~\ref{fig:modality}(b) reports per-stage F1 for each modality. scEEG is the stronger
  stand-alone modality in most stages, yet adding PPG improves fusion F1 in \emph{every} stage, most where
  cardiovascular dynamics are informative: $+12.2$ (REM), $+7.8$ (light), $+7.0$ (wake) over
  scEEG; deep sleep, dominated by slow waves, changes little (PPG alone $28.1\%$ F1; $+1.4$).
  PPG is thus valuable not as a stronger stand-alone signal but as a complementary one
  supplying autonomic cues scEEG lacks.}

\subsection{Comparison of Fusion Strategies}
The progressive improvement from score-level ($\kappa=0.747$) to cross-attention ($\kappa=0.767$) to Mamba-enhanced fusion ($\kappa=0.798$) reveals two key insights. First, the improvement from score-level to cross-attention demonstrates that learned feature-level interactions capture complementary information that simple probability averaging cannot exploit. Second, \rev{although the temporal modules perform comparably, a parameter-matched ablation (Table~\ref{tab:tcm_ablation}) shows that
Bi-Mamba attains the highest per-subject $\kappa$, matching the Transformer
and edging out the gated RNNs, so we adopt it for both this marginal
accuracy edge and its linear-time efficiency, a consideration that is also
central to long-sequence temporal modelling more
broadly~\cite{deng2026dynamic}, while the overall improvements are driven
primarily by the multimodal fusion rather than by a specific temporal
architecture.}

\subsection{Implications for Lightweight Sleep Monitoring}
Prior validation studies of consumer sleep trackers have commonly adopted acceptable thresholds of $\leq$30\,min for TST and $\leq$5\% for SE when compared to polysomnography~\cite{schyvens2025performance}. Our Mamba-enhanced fusion model substantially exceeds these benchmarks, achieving TST MAE of 13.78\,min (54\% below the 30\,min threshold) and SE MAE of 2.38\% (52\% below the 5\% threshold). These results suggest that scEEG-PPG fusion may offer a promising approach for sleep architecture assessment\rev{,  motivating future validation on} lightweight devices, potentially supporting applications such as longitudinal sleep monitoring and treatment response evaluation.

\subsection{Limitations}
\label{sec:discussion}
Several limitations should be acknowledged. First, restricting the input to
single-channel scEEG and PPG inherently caps performance relative to
multi-channel or multi-signal devices, a deliberate trade-off between
clinical-grade accuracy and practical wearability. Second, we evaluated only
three fusion strategies; other approaches such as multi-task
learning~\cite{zhao2024sequence} and graph neural
networks~\cite{wang2022multi} remain for future work. Third, the scEEG
signals used in this study were extracted from traditional multi-channel PSG
systems (C4-M1 for MESA, C3-M2 for CFS and ABC); validation with lightweight
wearable EEG devices (e.g., forehead scEEG sensors as shown in
Fig.~\ref{fig:wearable}(d)) in naturalistic settings would better reflect
real-world performance. \rev{Fourth, the fine-tuned CFS results used a random subject-level split without family-aware control, so mild optimism from related individuals across splits cannot be excluded; the zero-shot results (Table~\ref{tab:fusion_cross_dataset}) involve no target-domain training and are free of this concern.} \rev{Fifth, our robustness analysis probes temporal misalignment but not other asynchronous-acquisition factors such as missing data and motion artefacts, which remain important for prospective on-device validation.}

\section{Conclusion}
This work investigates the extent to which PPG can complement scEEG for lightweight sleep staging. Our evaluation demonstrates that 3-minute windows provide an effective trade-off between classification performance and window length for both modalities \rev{under a short-window, retrospective-scoring setting}. Among three fusion strategies investigated, Mamba-enhanced fusion achieves the best performance ($\kappa=0.798$, $Acc=86.9\%$), with notably improved light sleep classification (F1-score: 85.63\% vs. 77.76\%, recall: 82.85\% vs. 69.95\% for scEEG alone). Cross-dataset evaluation on CFS and ABC confirms that learned fusion strategies generalize effectively across populations \rev{after light target-domain fine-tuning}. These findings suggest that combining scEEG and PPG as two \rev{lightweight} modalities is a promising approach to address scEEG's limitation in light sleep classification. This study provides evidence \rev{motivating the future} development of practical multimodal solutions, such as smart headbands integrating both sensors\rev{, which would require prospective on-device validation,} for accessible sleep health monitoring.

\section*{References}

\bibliographystyle{unsrt}
\bibliography{references}

\clearpage
\onecolumn



\end{document}